\documentclass[prx,preprint,english]{revtex4}
\usepackage{fancyhdr}
\usepackage{amsmath,amsfonts,amssymb}
\usepackage{graphicx}
\usepackage{color}
\usepackage{bbm}

\begin{document}
\title{Localization in the Kicked Ising Chain}
\author{Daniel Waltner, Petr Braun}
\address{Fakult\"at f\"ur Physik, Universit\"at Duisburg-Essen, Lotharstra\ss e 1, D-47048 Duisburg, Germany}
\begin{abstract}
Determining the border between ergodic and localized behavior is of central interest for interacting many-body 
systems. We consider here the recently very popular spin-chain model that is periodically excited.  A convenient description of such a many-body system is achieved by the dual operator that evolves
the system in contrast to the time-evolution operator not in time but in particle direction. We identify in this
paper the largest eigenvalue of  a function based on the dual operator as a convenient tool to determine if the system shows ergodic 
or many-body localized features. By perturbation theory in the vicinity of the noninteracting system we explain analytically the eigenvalue structure and compare it with numerics in [P.\ Braun, D.\ Waltner, M.\ Akila, B.\ Gutkin, T.\ Guhr, Phys.\ Rev.\ E {\bf101}, 052201 (2020)] for small times. Furthermore we identify a quantity that allows based on extensive large-time numerical computations of the spectral form factor to distinguish between localized and ergodic system features and to determine the Thouless time, i.e.\ the transition time between these regimes in the thermodynamic limit.    
\end{abstract}
\maketitle
\section{Introduction}

Interacting spin systems play a prominent role in the study of  many-body quantum
systems, in particular, of many-body quantum chaos. Prominent 
examples are the Ising model \cite{Bogomolny,Keating,Heidel,Moessner0,Papic},
the Heisenberg model \cite{Znidaric,Langer,Gemmer} but also other models like the Bose-Hubbard model
\cite{Urbina,Dubertrand} and coupled kicked tops \cite{And} are considered. In the experimentally  most
relevant case of spin $1/2$ semiclassical methods successful in the theory
of the few-body chaos \cite{Fritz,Stockmann} are inapplicable and alternative approaches
are needed. A very popular one applicable to a large class of kicked chains is the method
of the dual operator in which the evolution time $t$ and the particle number $N$ exchange
their roles \cite{Akila}; it is especially suited to study the thermodynamic
limit when $N$ is very large. Statistical properties of the spectra are then
expressed in terms of the \textquotedblleft dual operator\textquotedblright%
\ replacing the Floquet operator of the standard approach.

In the last years this dual perspective became extremely popular to describe spectral properties 
\cite{Prosen, Chalker,ChalkerI,Braun,Flak}, to quantify 
entanglement \cite{ProsenII,Pal,Lamacraft}, correlation 
functions \cite{Gutkin,ProsenIII,LamacraftI} and the connection between 
quantum and classical system properties \cite{AkilaI,ChalkerII}.
The dual operator is usually non-unitary, however in the case of the so
called self-duality or self-unitarity, it is. A remarkable property of
strongly disordered self-dual chains is that they allow on the one hand for an
analytically exact description of many system properties and on the other hand
they are completely chaotic, two features usually considered as contradictory.
This applies to the spectral form factor \cite{Prosen,Chalker,ChalkerI,Braun,Flak} determining the
correlations between discrete quantum levels that behaves in accordance with
Random Matrix Theory (RMT) \cite{Meta, Guhr} already for short times as long as the
number of constituents $N$ of the chain is large, to the entanglement entropy
\cite{ProsenII,Pal,Lamacraft} and correlation functions
\cite{Gutkin,ProsenIII,LamacraftI}. The RMT behavior of the disordered kicked
Ising chain (KIC) and ergodicity \cite{Prosen0} with parameters corresponding to self-duality have been
proven in \cite{Prosen}. The method was based on analytic averaging of the
form factor in the dual representation.

Whereas Ref.\ \cite{Prosen} concentrated on the self-dual parameter regime, in
Ref.\ \cite{Braun} we studied the spectral form factor of the strongly
disordered KIC in the regime where the self-duality condition is violated by
changing the Ising interaction constant. When that constant is brought to zero
the resulting chain of independent spins is trivially localized such that
somewhere on the way a transition between the ergodic and localized phase is
bound to take place. First indications of that transition were given in \cite{Braun}.
However, the drawback was here that the standard $N$ resolved quantities to detect
a transition to many-body localization like the spacing ratio or the entanglement entropy
considered can only be obtained for small $N$, see e.g. Ref.\ \cite{Papic}. An analogous statement applies to the 
largest eigenvalue of the double dual operator that is a function of the dual operator:
it can be computed only for a small number of timesteps \cite{Braun}. The aim of this paper is to solve
this problem: by extensive numerical calculations of the spectral form factor we are able
to compute this eigenvalue also for much larger times. This is especially relevant as  localization
 is a system feature that can only be probed in the limit of large (optimally infinite) times 
and system sizes. Independently, as the dual-operator method became during the last years an extremely popular method to analyse various system features \cite{Akila,Prosen,Chalker,ChalkerI,Braun,Flak,ProsenII,Pal,Lamacraft,Gutkin,ProsenIII,LamacraftI,AkilaI,ChalkerII, Prosenarx} the understanding of the eigenvalue structure of this operator is also a topical research question on its own.


The outline of the paper is as follows: In the next section we introduce the relevant 
quantities and recapitulate previous results.
In Section III we develop a perturbation theory starting from the 
case of noninteracting spins. Together with our analysis in the vicinity of self 
duality in Ref.\ \cite{Braun} we thus exploit all possible analytical
descriptions in the regime where the self-duality condition is violated. In
Section IV we briefly recapitulate our numerical calculations in Ref.\ \cite{Braun} for
the eigenvalue of the double dual operator of largest magnitude with
times $t$ up to $20$. In Section V we present a method to determine that
eigenvalue from the numerical form factor for times much larger than in the previous section but
smaller than the Heisenberg time $T_{H}$; we demonstrate its
characteristic change in the crossover between the ergodic and localized
regime. These results allow to determine the Thouless time as described in 
Section VI. We conclude in Section VII and present technical details in the Appendix.

\section{Basic Evolution Operators and the Spectral Form Factor}

We consider the disordered KIC with ring topology
consisting of $N$ spins $1/2$. The evolution operator of the system per
period, i.\ e. the Floquet operator, reads,%
\begin{align}
\hat{U}  & =\hat{U}_{I}\hat{U}_{b},\label{evolup}\\
\hat{U}_{I}  & =\exp\left(  -iJ\sum_{n=1}^{N}\hat{\sigma}_{n}^{z}\hat{\sigma
}_{n+1}^{z}\right)  ,\nonumber\\
\hat{U}_{b}  & =\exp\left(  -i\sum_{n=1}^{N}h_{n}\hat{\sigma}_{n}^{z}\right)
\exp\left(  -ib_{x}\sum_{n=1}^{N}\hat{\sigma}_{n}^{x}\right)  .\nonumber
\end{align}
The factor $\hat U_I$ describes the Ising interaction between the neighboring spins with the interaction constant   
$J$. The second factor in the operator $\hat{U}_{b}$ can be interpreted as a kick
by the magnetic field $b_x$ in the transverse direction while the first one
characterizes longitudinal local magnetic fields randomly distributed along
the chain. The operator $\hat{U}$ acts in the Hilbert space with the dimension
$2^{N}$ spanned by the basis vectors $\left\vert \sigma_{1},\ldots,\sigma
_{N}\right\rangle$ with $\sigma_{n}=\pm1$. We consider here as in Ref.\ \cite{Braun}
 the orthogonal kick
strength  fixed, $b_{x}=\pi/4$
and the Ising constant changing from $J=\pi/4$ to $0$. 

The spectral form factor is defined by,%
\begin{equation}\label{formfac}
K_{N}\left(  t\right)  =\left\langle \left\vert \operatorname*{Tr}\hat{U}%
^{t}\right\vert ^{2}\right\rangle /T_{H},
\end{equation}
where $T_{H}=2^{N}$ is the Heisenberg time for the chain of spins $1/2$;
averaging is over the disorder realizations. If the disorder on different spins is
uncorrelated averaging can be performed analytically provided the disorder
statistics is given. Assuming the local fields $h_{n}$ to be Gaussian-distributed it was 
shown in Refs.\ \cite{Prosen,Braun} that
\[
K_{N}\left(  t\right)  =\operatorname*{Tr}\left(  \mathcal{A}_{\xi}\left(
t\right)  \right)  ^{N}/T_{H}.%
\]
The time-dependent operator $\mathcal{A}_{\xi}(t)$ acts in the "squared" dual Hilbert space with the
basis $\left\vert \pmb{\sigma\sigma}^{\prime}\right\rangle =\left\vert
\pmb{\sigma}^{t}\right\rangle \otimes\left\vert \pmb{\sigma}^{\prime
t}\right\rangle $ and dimension $2^{2t}.$ It is given by the Kronecker product%
\[
\mathcal{A}_{\xi}(t)=\left(  \overline{\hat{w}\left(  t\right)  }\otimes
\overline{\hat{w}^{\ast}\left(  t\right)  }\right)  \mathcal{O}_{\xi},%
\]
where $\overline{\hat{w}(t)}$ is obtained from the factors of the dual operator 
$\hat{W}_n$ defined in Ref.\ \cite{Braun} by replacing  $h_{n}$ by its
disorder-averaged value $h=\left\langle h_{n}\right\rangle $; the operator
$\mathcal{O}_{\xi}$ depends on the disorder strength $\xi$. 
In the limit of large $\xi$ the operator $\mathcal{A}_\xi(t)$ turns to the double dual operator%
\begin{equation}
\mathcal{A}(t)=\mathcal{P}\left(  \overline{\hat{w}\left(  t\right)  }\otimes\overline
{\hat{w}^{\ast}\left(  t\right)  }\right)  \mathcal{P}.\label{ourA}%
\end{equation}
The operator $\mathcal{P}$ is a projector diagonal in the squared dual space
with the non-zero elements%
\[
\left\langle \pmb{\sigma\sigma}^{\prime}\left\vert \mathcal{P}\right\vert
\pmb{\sigma\sigma}^{\prime}\right\rangle =\left\{
\begin{array}
[c]{c}%
1,\quad\sum_{k=1}^{t}\sigma_{k}=\sum_{k=1}^{t}\sigma_{k}^{\prime},\\
0,\quad\text{otherwise.}%
\end{array}
\right.
\]
Since $K_{N}\left(  t\right)  =\sum_s\lambda
_{s}^{N}\left(  t\right)  /2^{N}$ where $\lambda_{s}(t)$ are the eigenvalues of
$\mathcal{A}(t)$, we found in Ref.\ \cite{Braun} that only  the contribution of the largest $\lambda_{s}(t)$ by magnitude denoted by $\lambda_0(t)$ is of importance that turns out to be real and positive survives when
$N\rightarrow\infty$. As proved in \cite{Prosen} in the self-dual case $\mathcal{A}(t)
$ has $2t$ eigenvalues equal to one \cite{footn}, all other
eigenvalues having $\left\vert \lambda_{s}\right\vert <1$. Therefore the form
factor in the thermodynamic limit tends to $2t/T_{H}$ which is the result 
obtained by the Gaussian Orthogonal Ensembe (GOE) of RMT
in the limit of times much smaller than the Heisenberg
time. 



For $t$ prime and $\Delta J\equiv\pi/4-J$ small we obtained for the largest eigenvalue
of the double dual operator denoted by $\lambda_0(t)$ within perturbation theory \cite{Braun}%
\begin{equation}
\lambda_{0}\left(  t\right)  =1+\Delta J^{2}\frac{2t\left(  t-1\right)
}{2^{\left(  t-1\right)  /2}+1}+O\left(  \Delta J^{4}\right)
;\label{PTsmalldJ}%
\end{equation}
all other eigenvalues of the multiplets shift from one downwards thus forming
a gap between $\lambda_{0}\left(  t\right)  $ and the other levels for nonzero $\Delta J$. 
The relation (\ref{PTsmalldJ}) with $\Delta
J$ fixed and time growing implies that the gap diminishes  
proportional to $2^{-t/2}$ with the
ergodic form factor restored. 

\section{Perturbation theory starting from non-interacting spins}

In the other
extreme of non-interacting spins, $J=0$ resp. $\Delta J=\pi/4$ when the
evolution operator is a product of rotations of individual spins,
\begin{align*}
\hat{U}  & =\prod_{n=1}^{N}\hat{u}_{n},\\
& \hat{u}_{n}=\exp\left(  ih_{n}\hat{\sigma}_{n}^{ z  }\right)
\exp\left(  i\frac{\pi}{4}\hat{\sigma}_{n}^{x}\right)
=\exp\left(  i\gamma(h_{n}) \mathbf{e}_{n}\pmb{\sigma}_n/2  \right)  .
\end{align*}
The rotation around the $x-$direction by $\pi/2$ followed by the random
rotation by $2h_{n}$ about $z$ can be replaced by a single rotation about some
axis $\mathbf{e}_{n}$ by the angle $\gamma_{n}=\gamma\left(  h_{n}\right)  $
connected with $h_{n}$ by  $1+2\cos2\gamma_{n}=\cos2h_{n}$ or
\begin{equation}\label{gamma}
\gamma(h)=\arccos\frac{\cos2h-1}{2}.
\end{equation}
Averaging over
disorder can be performed independently for each spin, and we get, for all
$N$,%
\[
K_{N}\left(  t\right)  =\left\langle \frac{1}{2}\left\vert \operatorname*{Tr}%
\hat{u}^{t}\right\vert ^{2}\right\rangle ^{N},
\]
where $\hat{u}$ is obtained from $\hat{u}_n$ by replacing $h_n$ by its disorder 
averaged value $h=\langle h_n\rangle$. 
Remembering $K_{N}\left(  t\right)  =\sum_s\lambda_{s}^{N}\left(  t\right)
/2^{N}$ we see that only one eigenvalue $\lambda_{0}\left(  t\right)  $ of the
operator $\mathcal{A}(t)$ is non-zero and equal to,%
\[
\lambda_{0}\left(  t\right)  =\left\langle \left\vert \operatorname*{Tr}%
\hat{u}^{t}\right\vert ^{2}\right\rangle ,
\]
i.\ e., the operator is proportional to a projector on a single vector in the
squared dual Hilbert space. This vector has highest possible symmetry; it is invariant under 
permutation and reflection of the individual spins as we show in appendix \ref{appA}. In the regime of 
small $\Delta J$ we found in Ref.\ \cite{Braun} that the dominant eigenvalue for small $t$  possessed the
same symmetry. 

It is easy to calculate $\lambda_{0}\left(  t\right)  $ in the case of strong
disorder assuming all angles of the $z-$rotation $2h_{n}$ to be uniformly distributed
in $[0,2\pi]$; then we obtain,%
\begin{align}
\lambda_{0}\left(  t\right)    & =\frac{4}{2\pi}\int_{0}^{2\pi}\cos^{2}%
t\gamma\left(  h\right)  dh\label{noninteract}\\
& =2+\left(  -1\right)  ^{k}\frac{\left(  2m-1\right)  !!}{\left(  2m\right)
!!},\quad k=\left[  \frac{t+1}{2}\right]  ,\quad m=\left[  \frac{t}{2}\right]
.\nonumber
\end{align}
In the limit of large times  the stationary phase approximation taking into account the stationary points $h=0$ and $h=\pi$ gives,%
\begin{equation}\label{lamunp}
\lambda_{0}\left(  t\right)  =2+\left(  -1\right)  ^{k}\sqrt{\frac{2}{\pi t}}.
\end{equation}
\begin{figure}[h]
 \begin{center}
  \includegraphics[width=10cm]{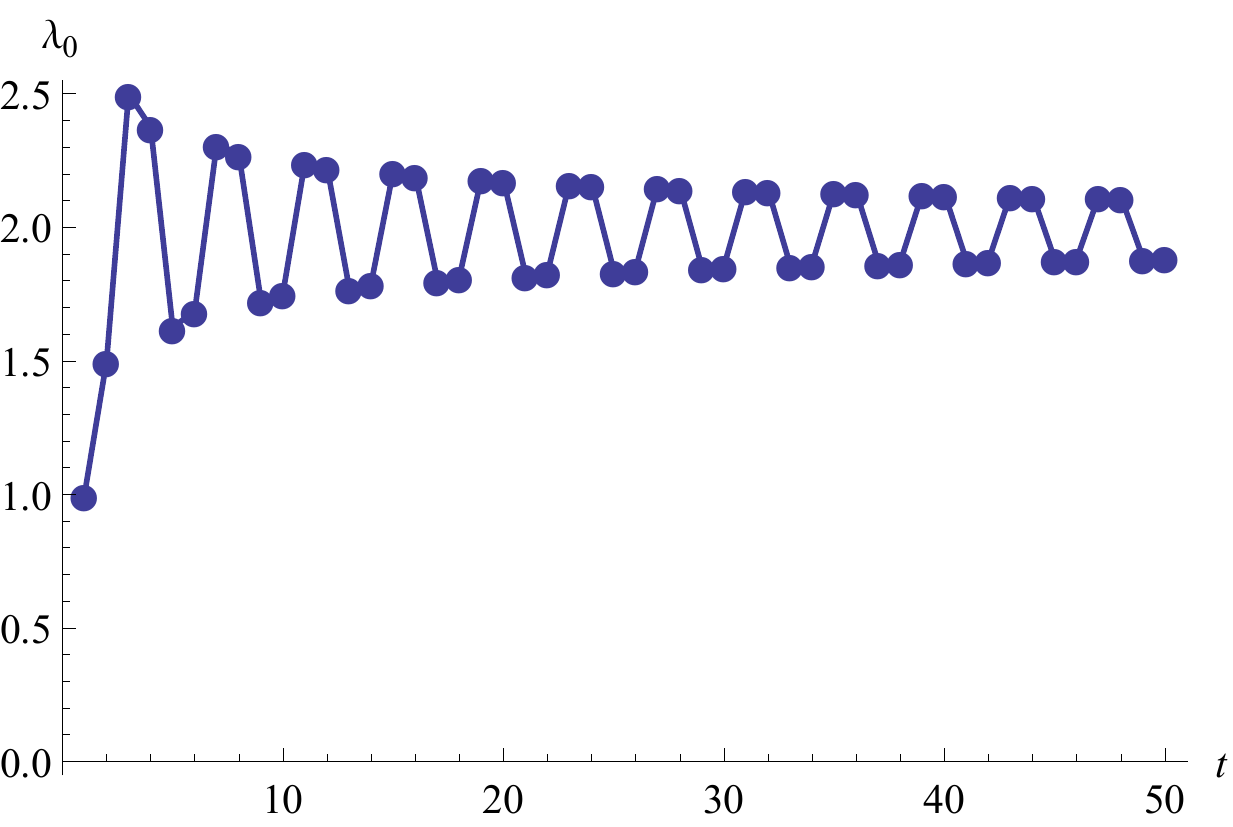}
\caption{Senior eigenvalue of $\mathcal{A}(t)$ against time, non-interacting spins
}
  \label{Fig3}
 \end{center}
\end{figure}
Approximate periodicity of $\lambda_{0}\left(  t\right)  $ with period $\Delta
T=4$ is associated with the $x-$rotation angle $\pi/2$; that can be
interpreted as the result of interference of the saddle-point contributions
to the integral (\ref{noninteract}). The oscillation with period  four die out $\sim t^{-1/2}$ with the
growth of time and are suppressed if we average $\lambda_{0}(t)$ over four
consecutive times; in both cases we get $\lambda_{0}=2$ and $K_N\left(  t\right)
=1$ which corresponds to the Poissonian RMT statistics.

\begin{figure}\begin{center}
\includegraphics[width=7.5cm]{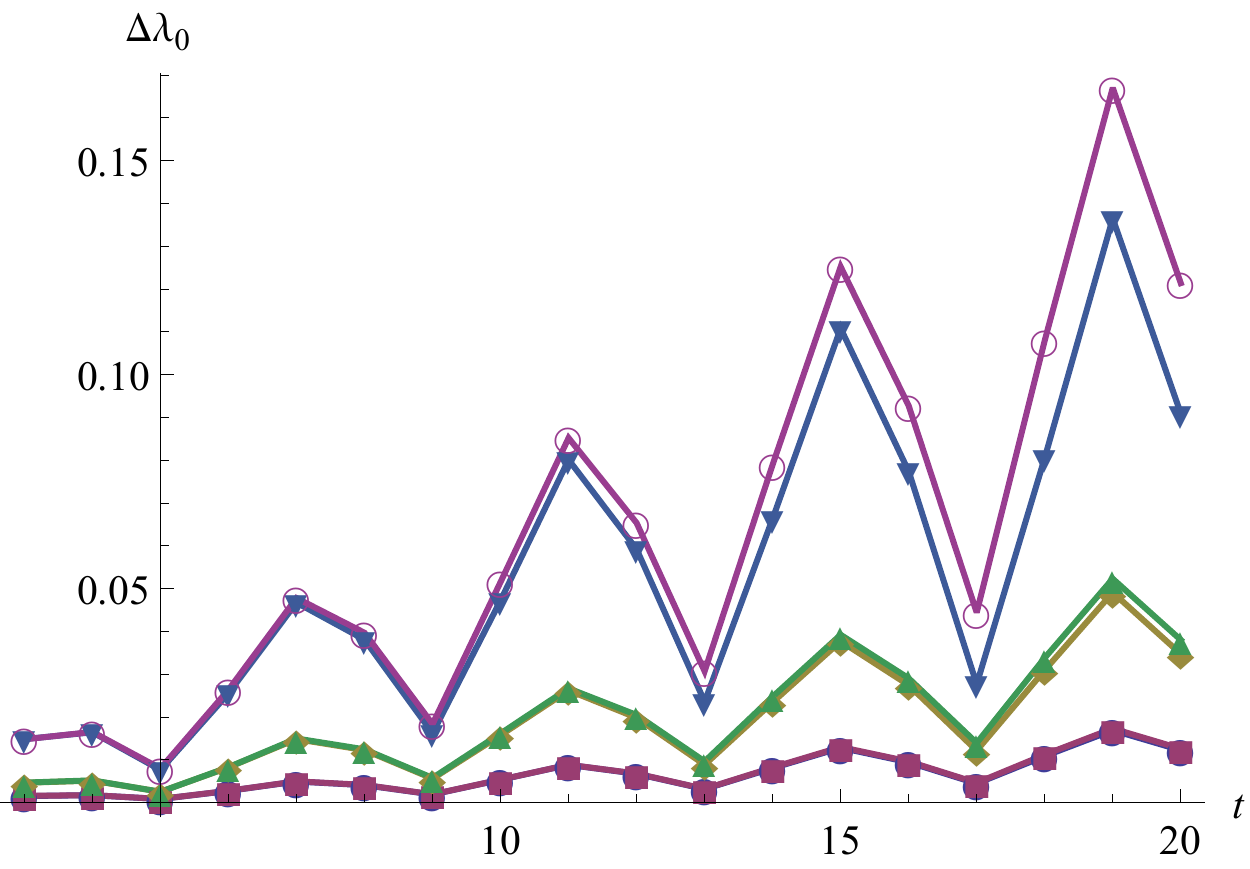}\hspace*{8mm}\includegraphics[width=7.5cm]{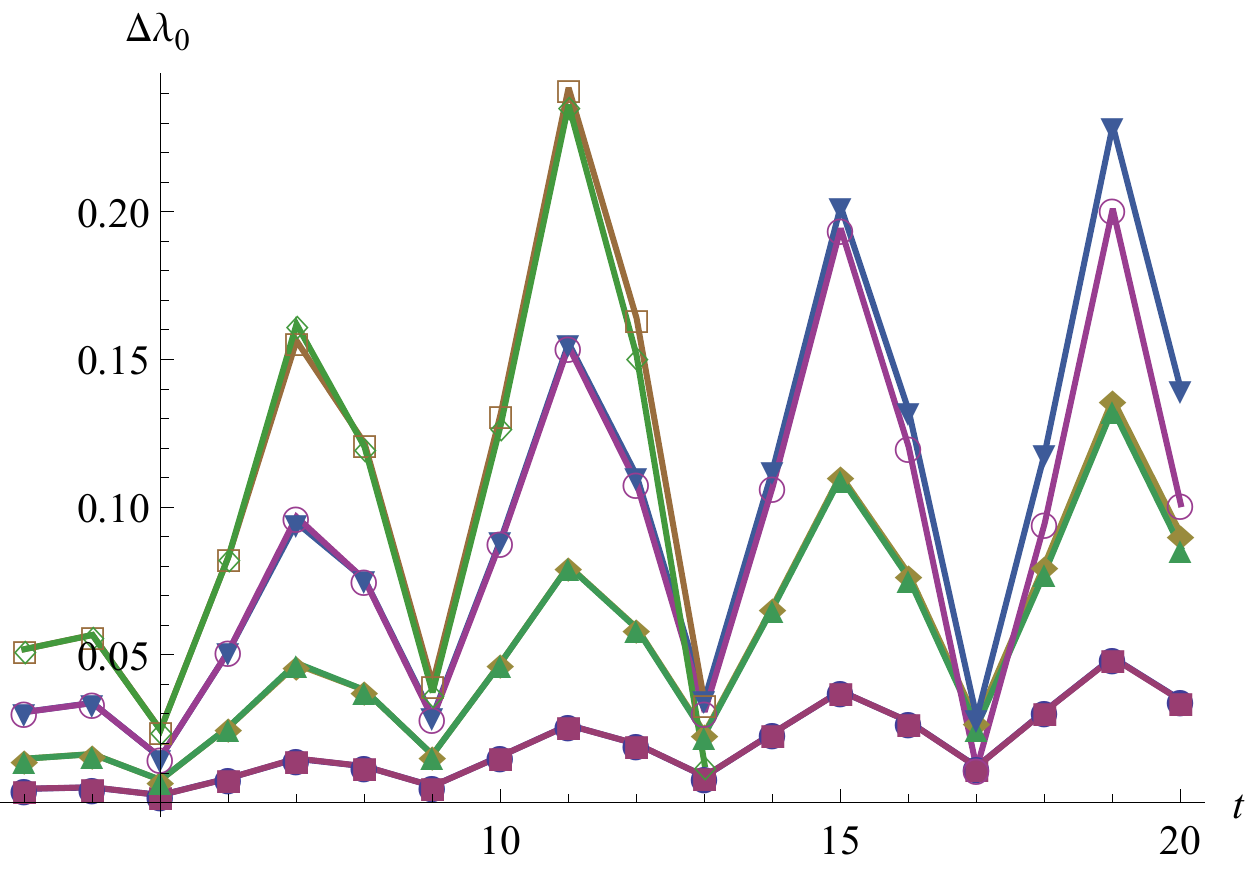}
\caption{Difference $\Delta\lambda_0(J)=\lambda_0(J)-\lambda_0(0)$ as a function of $t$. Left panel: Perturbation theory up to second order, different curve pairs for $\Delta J=0.8,\, 0.76,\, 0.74$ from bottom to top, for each $\Delta J$ the lower of the nearby curve
obtained numerically the upper one from perturbation theory.   Right panel: Perturbation theory up to fourth order, different curve pairs for $\Delta J=0.76,\,0.74,\,0.72,\,0.7$, upper of nearby curve obtained numerically, the lower one from perturbation theory.} \label{Fig8}
\end{center}\end{figure}

Perturbation theory can be used to get the eigenvalue $\lambda_{0}\left(
t\right)$ for $J$ non-zero but small. The deflection from the limit
(\ref{noninteract}) is quadratic in $J$; the method and the details of the calculation are described
in appendix \ref{appB}. Similar to the previous case, the radius of convergence
shrinks with the growth of time.
Here we get 
\begin{eqnarray}\label{final}
\lambda\left(  t\right)&=&\lambda_0\left(  t\right)+\frac{tJ^2}{\lambda_0\left(  t\right)}\left[t\left(2-\sum_{k=0}^{[t/2]}\left(\begin{array}{c}1/2\\k\end{array}\right)\right)^2-\lambda_0^2(t)\right.\\&&\left.-\sum_{\tau=1}^{t-1}\left(2\left(1-S(\tau-1)+S(t)-S(t-\tau-1)\right)-\left(\begin{array}{c}-1/2\\{[t/2]}\end{array}\right)\right)^2\right]\nonumber
\end{eqnarray}
with 
\begin{equation}
S(\tau)=\sum_{k=0}^{[\tau/2]}\left(\begin{array}{c}-1/2\\k\end{array}\right).
\end{equation}
In the left panel of Fig.\ \ref{Fig8} we compare this perturbative result with the the exact one from numerical
calculations. We confirm by this analytically the structure that $\lambda_0(t)$ for small nonzero $J$ oscillates around 
$\lambda_0(t)$ for $J=0$ found numerically in Ref.\ \cite{Braun}.
Considering the average of $\lambda_0(t)$ over the oscillations of period four 
\begin{equation}\label{lambdaav}
\langle\lambda_0(t)\rangle=\frac{1}{4}\left[\lambda_0(t+3)+\lambda_0(t+2)+\lambda_0(t+1)+\lambda_0(t)\right]
\end{equation}
we get as shown in appendix \ref{appB} in the limit of large time up to quadratic order in $J$ 
\begin{eqnarray}\label{final1}
\langle\lambda_0(t)\rangle&=&\lambda_0(t)+\frac{tJ^2}{\lambda_0(t)}\left[-\frac{4(\sqrt{2}-1)}{\sqrt{\pi}}\left[(2-\sqrt{2})\zeta[1/2]-\zeta[1/2,1/4]+\zeta[1/2,3/4]\right]\right.\nonumber\\&&\left.+(2-\sqrt{2})^2-4-\frac{4}{\pi}(\gamma+\ln t)\right]+O(J^4)
\end{eqnarray}
with the Euler constant $\gamma$, the Hurwitz zeta function $\zeta[a,s]$ and the Riemann zeta $\zeta[s]$. Note that the sum of the $t$-independent terms in the square bracket above are negative and about twice as large as the prefactor of the $\ln t$-term   implying that $\langle\lambda_0(t)\rangle$ decreases approximately linearly as a function of time. 
\begin{figure}\begin{center}
\includegraphics[width=10cm]{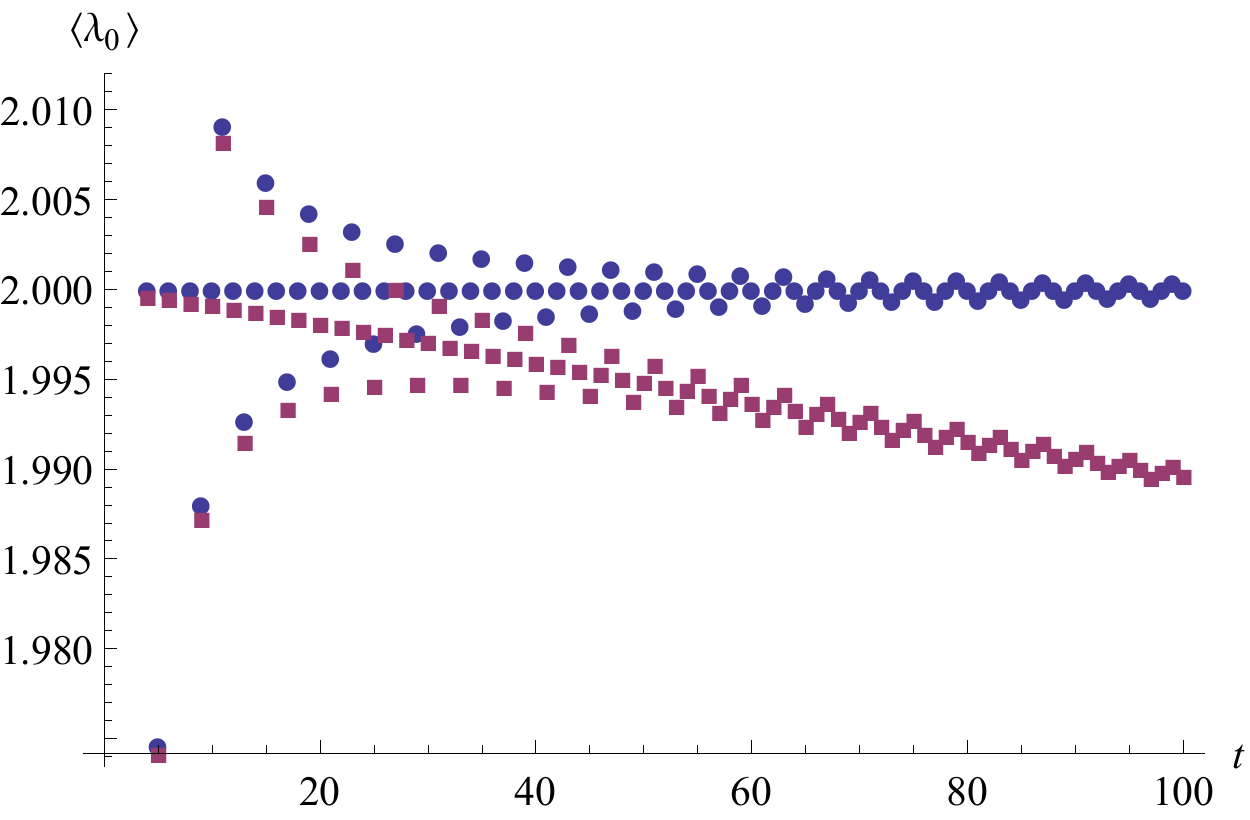}
\caption{$\langle\lambda_0(t)\rangle$ obtained from fourth order perturbation theory for $J=0$ (blue) and $J=\pi/4-0.78$ (magenta).}
\label{perlong}
\end{center}\end{figure}
Extending this perturbation theory to  higher orders, the third order
in $J$ vanishes due to similar reasons like the first order (see appendix \ref{appB}) and the fourth order can be computed in a similar way as the second order. However, as  the result is 
quite lengthy, we show the resulting plots in the right panel of Fig.\ \ref{Fig8} but we don't give its explicit form in appendix \ref{appB}. The full expression is contained in a mathematica file that can be obtained from the authors upon request. The functional form of $\langle\lambda_0(t)\rangle$ predicted after Eq.\ (\ref{final1}) is also confirmed in Fig.\ \ref{perlong} where we show $\langle\lambda_0(t)\rangle$ computed up to fourth-order perturbation theory. We here confirmed the 
validity of the perturbation theory by checking that the fourth order contribution is much smaller than the second order one. 

\section{Direct numerical calculations}



\begin{figure}[h]
 \begin{center}
  \includegraphics[width=10cm]{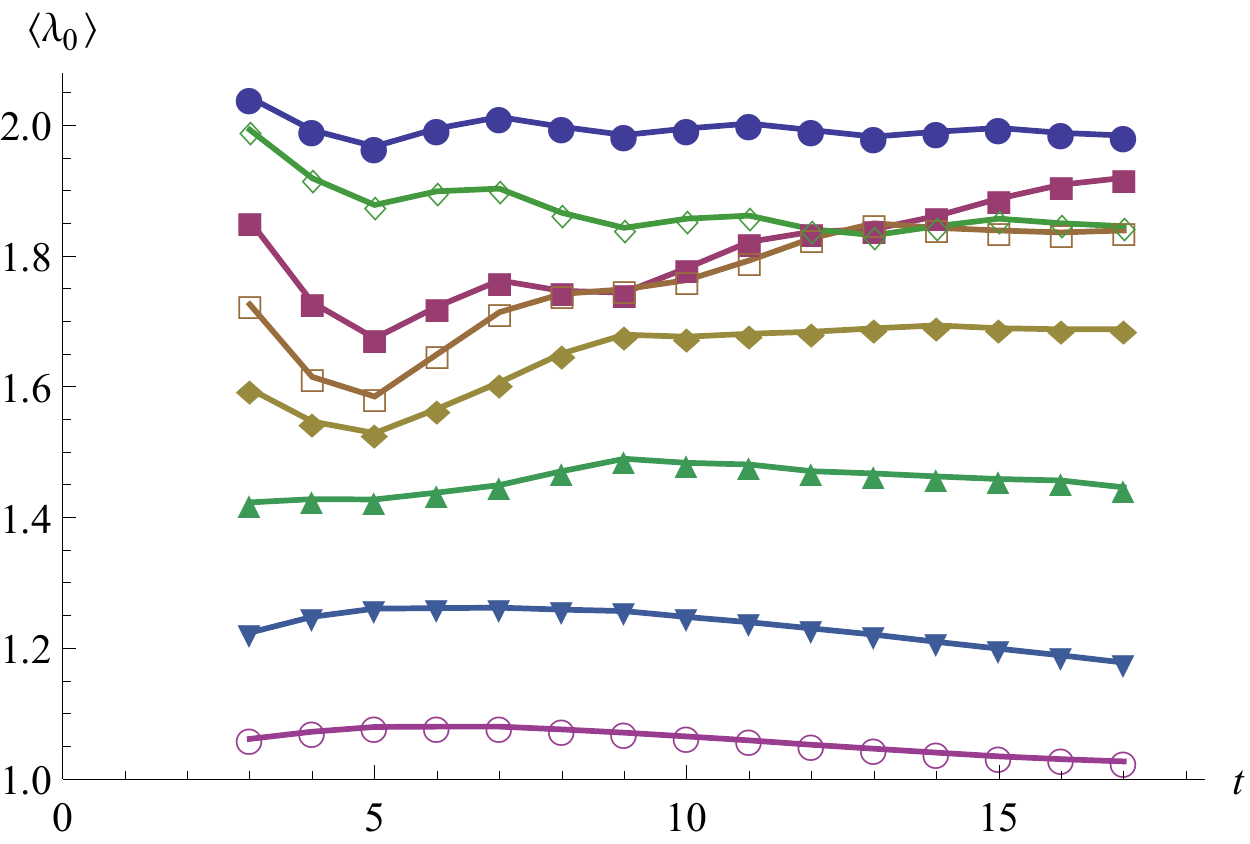}
\caption{Senior eigenvalue smoothed over the period 4 of the system-specific oscillations for $\Delta J =0.1,\,0.2,\,0.3,\,0.4,\,0.5,\, 0.6,\, 0.7,\,\pi/4$ in the order upwards at the left border of the plot.
}
  \label{Fig2}
 \end{center}
\end{figure}

For $t\leq20$ the eigenvalue
$\lambda_{0}\left(  t\right)  $ was calculated in Ref.\ \cite{Braun}. 
The results there are marred by the non-universal oscillations in time with
period four. To suppress them, we average here over an interval of four consecutive times
as described in Eq.\ (\ref{lambdaav});
the averaging bracket will be dropped in the future. The eigenvalue plot
obtained after the smoothing is shown in Fig.\ \ref{Fig2}.

One sees that the two lowest curves ($\Delta J=0.1$ and $0.2)$ tend to go down
such that ergodic behavior can be expected at larger times. The uppermost
curve, practically constant at the level $\lambda=2,$ corresponds to the
trivially localized case of non-interacting spins. As regards to other values of
$\Delta J$ the large-time behavior predictions are not possible from these
data. What strikes the eye is the irregularity of the behavior of the curves with
$\Delta J=0.5,0.6,0.7$; they cross and change their relative positions at the
right border of the plot compared with its left border. By different methods in
\cite{Braun} a transition of the KIC from ergodicity to localization occurring at
$\Delta J=0.58$ was found; below we shall see a direct evidence of that transition.

%
%

\section{Large times, all $J$: indirect estimate of $\lambda_0$}

By "large" we mean here times of the order of the Heisenberg time
$T_{H}=2^{N}$. 
We find it numerically most effective to estimate $\lambda
_{0}\left(  t\right)  $ from the numerically calculated spectral form factor
for several consecutive chain lengths $N$. It can be done if $\lambda
_{0}\left(  t\right)  $ is separated by a gap from all others; we found in Ref.\ \cite{Braun} that this is
true at least in the limits of weak interaction and for small times for all
non-zero $\Delta J$.   
Then for sufficiently large $N$ the form factor will be
dominated by that single eigenvalue of the double dual operator growing
exponentially with $N,$
\begin{equation}
K_{N}\left(  t\right)  \approx\left[  \lambda_{0}\left(  t\right)  /2\right]
^{N}\label{Kaspowla}%
\end{equation}
which corresponds to the localized regime. We showed in Ref.\ \cite{Braun} that such a behavior follows
if the system splits into independent subsystems. The  gap between $\lambda
_{0}\left(  t\right)  $ and the other eigenvalues of the double dual operator depends on
time; if it shrinks to zero that estimate ceases to be true. On the other hand, in the ergodic
limit the form factor obeys the GOE prediction, $K_{N}^{\rm{GOE}}(t)=2t/2^{N}-2\left(
t/2^{N}\right)  ^{2}+\ldots$ ; at large $N$ the form factor is then
practically $N-$independent apart from the trivial scaling with $T_{H.}=2^{N}$.

We now calculate the ratio,%
\begin{equation}\label{lamratio}
\lambda_{0,N}\left(  t\right)  =\frac{2K_{N}\left(  t\right)  }{K_{N-1}\left(
t\right)  }%
\end{equation}
for sufficiently large $N$. If the chain is in the localized domain and
(\ref{Kaspowla}) is applicable the result must be close to the $N-$independent
$\lambda_{0}\left(  t\right)  $.  That can be checked by comparison of the
empirical $\lambda_{0,N}\left(  t\right)  $ and $\lambda_{0,N-1}\left(
t\right)  $; if these are not too different we achieved convergence in $N$ and
can assume,%
\[
\lambda_{0}\left(  t\right)  \approx\lambda_{0,N}\left(  t\right)
\approx\lambda_{0,N-1}\left(  t\right)  ,
\]
with the accuracy increasing with  $N$.

\begin{figure}[h]
 \begin{center}
  \includegraphics[width=10cm]{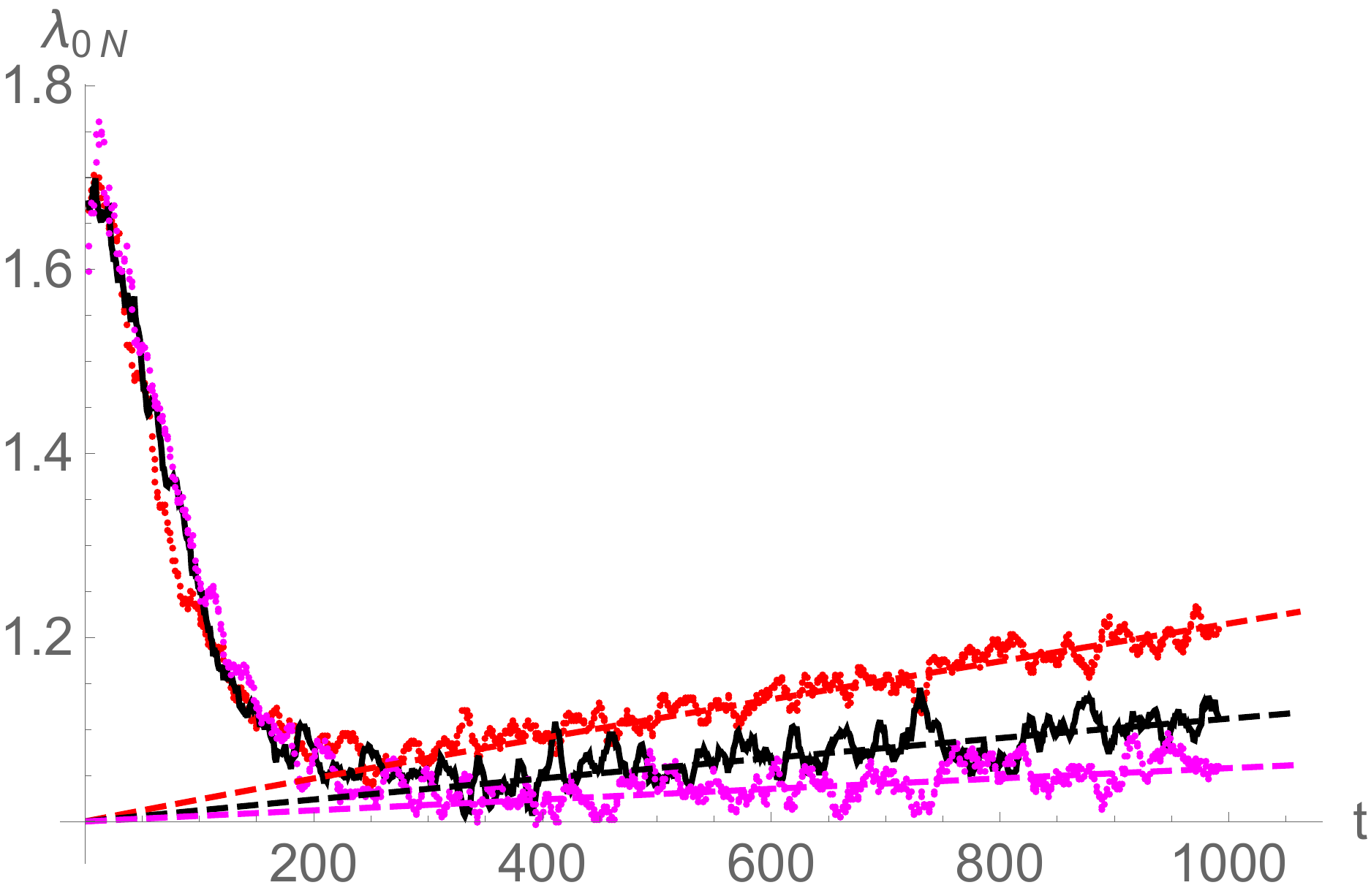}
\caption{Approximate $\lambda_{0,N}(t)$ against time obtained as ratio of form factors of chains with  12 to 11 spins (red), 13 to 12 spins (black) and 14 to 13 spins (magenta), case $\Delta J=0.4$  
with the corresponding GOE predictions (dashed lines).
}
  \label{Fig5}
 \end{center}
\end{figure}

We used that approach  calculating the form factor of  the KIC with $N=11-15$;
the averaging was performed over 1000 disorder realizations. Consider, e.\ g. the result for
$\Delta J=0.4$ in Fig. \ref{Fig5} where $\lambda_{0,12}\left(  t\right)  $ (red),  $\lambda_{0,13}\left(  t\right)  $ (black) 
and $\lambda_{0,14}\left(  t\right)$ (magenta) are depicted.  The three
dashed straight lines passing through the origin indicate that  $\lambda_{0,N}(t)$ calculated
for spectra obeys the GOE statistics,%
\begin{equation}
\lambda_{0,N}^{\mathrm{GOE}}=\frac{2K_{N}^{\mathrm {GOE}}\left(  t\right)  }{K_{N-1}^{\mathrm{GOE}}\left(
t\right)  }=1+\frac{t}{2^{N}}+\ldots\label{GOElambda}%
\end{equation}
for $N=12,13,14$. All three curves first go down following, apart from a small
initial stretch, an almost straight line practically the same for all $N$ considered.  Therefore they provide a sufficiently accurate approximation for the
eigenvalue $\lambda_{0}\left(  t\right)$ in that interval of times.  At
larger times  the three curves diverge approaching their "own" GOE lines: they
first  curve up and later merge with $\lambda_{0,N}^{\mathrm{GOE}}$ signaling the
transition from the localized regime to ergodicity.

The overall plot of $\lambda_{0,N}(t)$ with two almost straight stretches at the
ends  reminds of a hyperbola; in fact fitting them with a hyperbola one of whose
asymptotes coincides with $\lambda_{0,N}^{\rm{GOE}}$ proved to be fairly accurate, see appendix \ref{hyperbola}.

The GOE limit line approaches $\lambda=1$ with the increase of $N$.
Extrapolating to $N\rightarrow\infty$ we can thus conjecture that the exact
$\lambda_{0}\left(  t\right)  $ consists of the straight line with the
negative slope insignificantly different from that of $\lambda_{0,12/13/14}\left(
t\right)$.  
If we neglect the transition zone between the two regimes the overall plot of $\lambda
_{0}\left(  t\right)  $  thus predicted  consists of two straight lines and can
be viewed as a degenerate hyperbola.

\begin{figure}[h]
 \begin{center}
  \includegraphics[width=10cm]{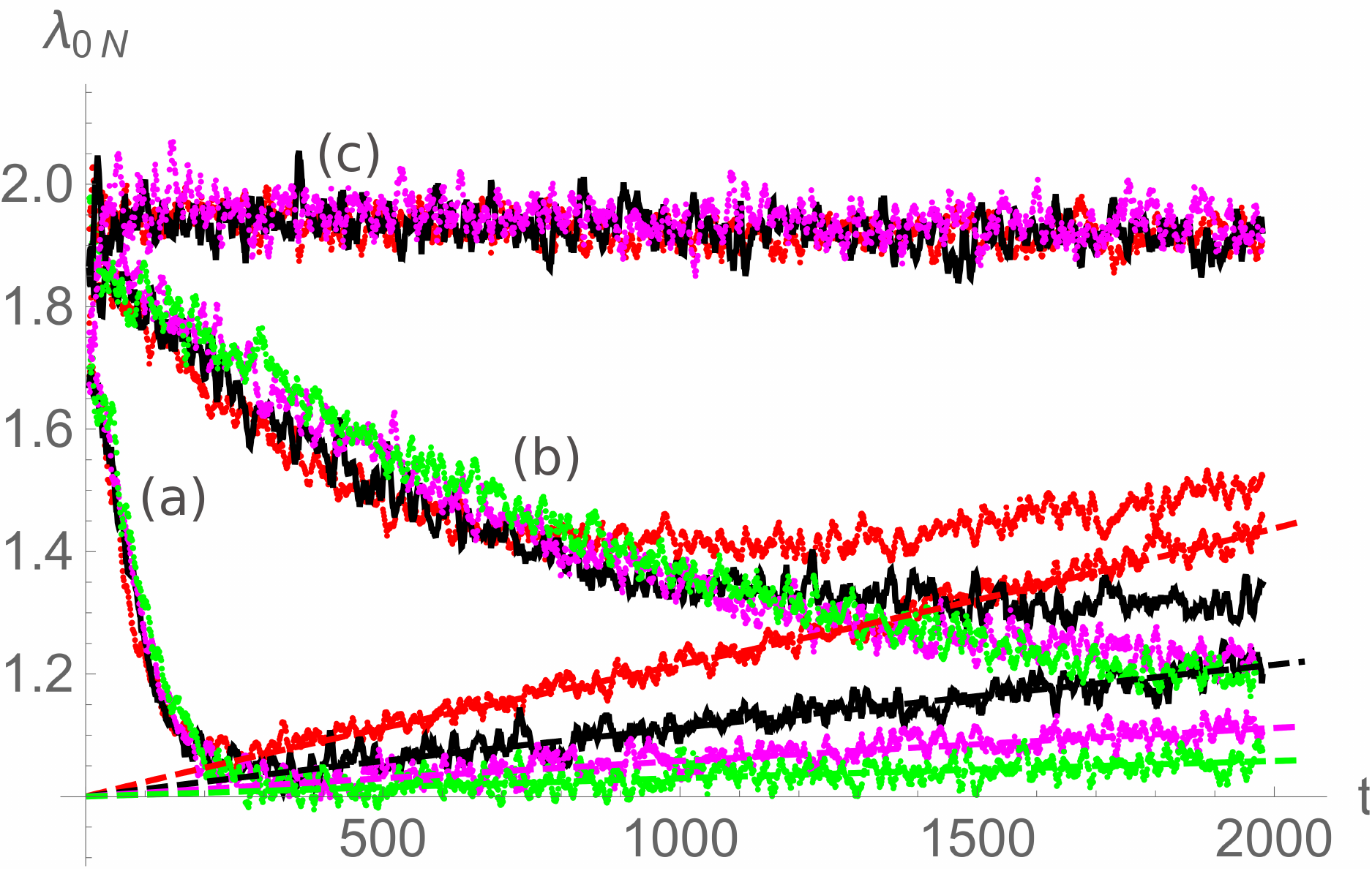}
\caption{Plots of $\lambda_{0,N}(t)$ against time  for $\Delta J=0.4 (a), 0.5 (b) , 0.6 (c)$ with $N=12$ (red), $13$ (black), $14$ (magenta) and $15$ (green). For a given $\Delta J$  the descending parts of the curves with different $N$  practically coincide  indicating convergence in $N$. For larger times the plots tend to the respective GOE limits. For $\Delta J=0.6$ (beyond the localization threshold) the plots are horizontal
}
  \label{Fig7}
 \end{center}
\end{figure}

The family of similar plots  of $\lambda_{0,N}(t)$ for several values of $\Delta J$ is shown in Fig.\ \ref{Fig7}, with $a, b, c$ standing for $\Delta J=0.4,0.5,0.6$, respectively and the color indicating the value of  $N$.  In
all cases there is good agreement between $\lambda_{0,N}(t)$ with $N=12-15$ at the descending stretch; at larger times the
curves merge with the respective $\lambda_{0,N}^{\mathrm{GOE}},\quad N=12-15$. The negative slope of the decreasing stretch tends after a decay for small $t$ (see Figs.\ \ref{Fig8},\ref{perlong}) to zero with the
growth of $\Delta J$ and disappears for $\Delta J\geq0.6$ .%

\begin{figure}[h]
 \begin{center}
  \includegraphics[width=10cm]{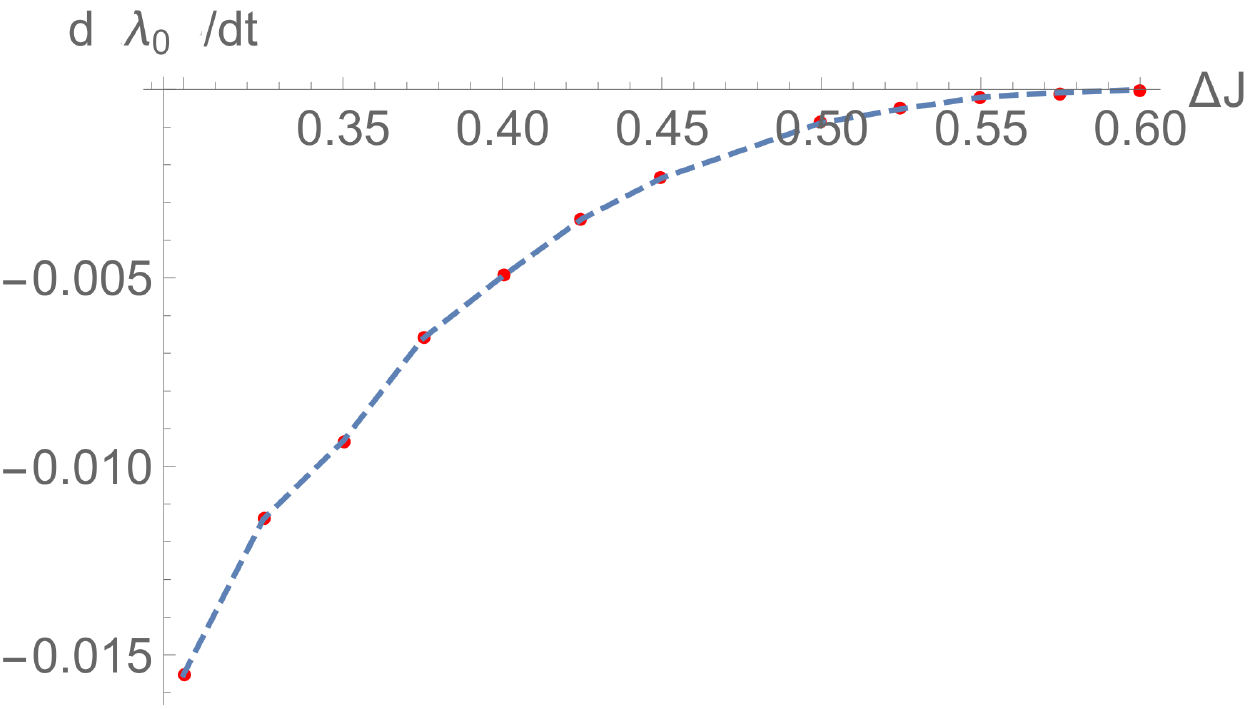}
\caption{Slope of the linear stretch of $\lambda_0(t)$ against $\Delta J$
}
  \label{Fig4}
 \end{center}
\end{figure}

 A more detailed analysis shows that  zeroing of the slope occurs
at the critical value $\Delta J=\Delta J_c\approx0.58$. It agrees with our previous finding for the
localization threshold in the KIC based on the spacing ratio statistics and the
entanglement entropy \cite{Braun}.
Therefore we determined the slope of the initial linear stretch of $\lambda
_{0,13}\left(  t\right)  $ as function of $\Delta J$ by fitting $\lambda
_{0,13}\left(  t\right)  $ by hyperbolas (see appendix \ref{hyperbola}) and calculated
$d\lambda_{0}/dt$ at $t=0.$ The results are shown in Fig. \ref{Fig4}.
The curve smoothly approaches zero at $\Delta J\approx0.58$ and stays zero at
larger $\Delta J$ which reminds of the behavior in the vicinity of a critical
point.

\section{Estimate of the Thouless time}

We want to use the results from the last section to give an estimate of the Thouless 
time of the system. It determines the time after that the system properties -- here the spectral form factor -- are described by RMT for large system size \cite{Thouless,Santos}.
We use in this context the formula smoothly interpolating between the localized and the RMT
regimes,%
\begin{equation}
K_{N}\left(  t\right)  \approx K_{N}^{\rm{GOE}}\left(  t\right)  +\left[  \lambda
_{0}\left(  t\right)  ^{N}-1\right]  /2^{N}.\label{Goeplusla}%
\end{equation}
It follows then that%
\begin{equation}
\lambda_{0}\left(  t\right)  \approx2\left[  K_{N}\left(  t\right)
-K_{N}^{\rm{GOE}}\left(  t\right)  +\frac{1}{2^{N}}\right]  ^{1/N}.\label{lambdaviak}%
\end{equation}
In the limit $N\rightarrow\infty$, time fixed, the GOE form factor tends to
zero and we get the expression for $\lambda_{0}\left(  t\right)
=\lim_{N\rightarrow\infty}2\left[  K_{N}\left(  t\right)  \right]  ^{1/N}$ of
the localized regime; in the opposite limit of the ergodic regime we have
$K_{N}\left(  t\right)  =K_{N}^{\rm{GOE}}\left(  t\right)  $ and $\lambda_{0}=1.$

\begin{figure}
\includegraphics[width=10cm]{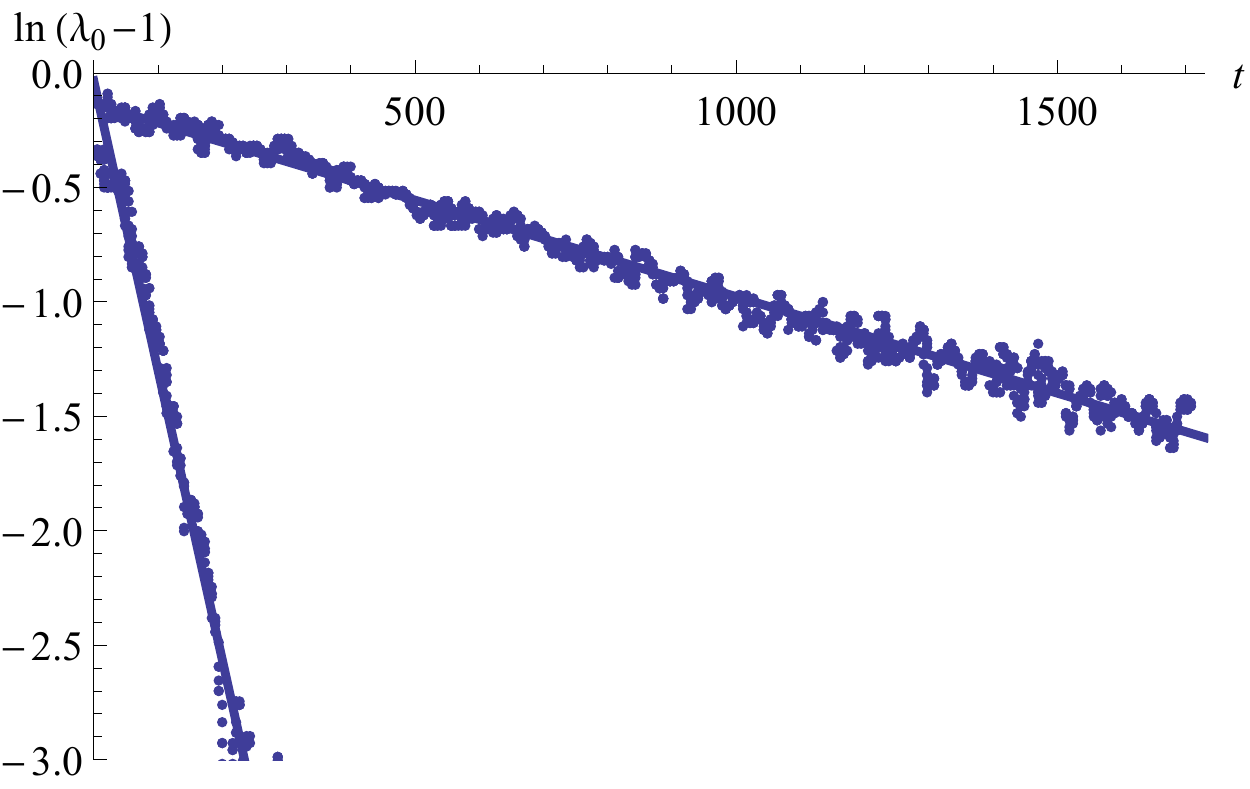}
\caption{Logarithmic plot of $\lambda_{0}-1$ with fit curves for $N=15$ and $\Delta J=0.4$ (left curve) and $\Delta J=0.5$ (right curve).}
\label{Fig10}
\end{figure}

The converged
$\lambda_{0}\left(  t\right)  $ decreases from its maximum $\lambda
_{m}=\lambda_{m}\left(  \Delta J\right)$, $1<\lambda_{m}<2,$ reached at
$t\sim10$ to its ergodic regime value $\lambda_{0}=1$.  In the regime of $\lambda_{0}\left(
t\right)$ not very close to $1$ an exponential fitting turns out to be
fairly accurate,
\begin{equation}
\lambda_{0}\left(  t\right)  =1+\left(  \lambda_{m}-1\right)  e^{-\eta
t},\label{expappro}%
\end{equation}
as shown in Fig.\ \ref{Fig10} for $N=15$ and  $\Delta J=0.4$, $\Delta J=0.5$. As in that $t$ interval
the corresponding curves for different  $N$ look very similar we don't show them in Fig.\ \ref{Fig10}.
The coefficient $\eta\left(  \Delta J\right)  $ decreases with the growth of
$\Delta J$ tending to zero at about $\Delta J=\Delta J_{c}\approx0.58$. 
 For $\Delta J_{c}<$ $\Delta J\leq\pi/4$ the fitted $\lambda_{0}$ is
time-independent increasing with the growth of $\Delta J\ $reaching $2$ in the
limit of non-interacting spins $\Delta J=\pi/4$.

The characteristic time $1/\eta$ of the decay of $\lambda_{0}\left(  t\right)
$ can be connected with the Thouless time $t_{Th}$ . Let us define it as the
time at which the deviation of the system form factor from the RMT prediction
$K_{N}^{\rm{GOE}}(t)$ becomes smaller than $K_{N}^{\rm{GOE}}(t)$ itself or, as follows from
(\ref{Goeplusla}) when $\ln t\sim N\ln\lambda_{0}\left(  t\right)  $.
Approximating $\ln\lambda_{0}\left(  t\right)  $ by $\left(  \lambda
_{m}-1\right)  e^{-\eta t}$ (see (\ref{expappro})) we obtain that the Thouless
time depends logarithmically on the system size,%
\begin{equation}
t_{Th}\sim\frac{1}{\eta}\ln N.
\end{equation}
The Thouless time determining the period until a quantity shows RMT features
naturally depends on the quantity and the system considered. Given that
Ref.\ \cite{Prosenarx} considers here a quantity different from the spectral form factor
explains the different ($N$ independent) scaling or the scaling with $N^2/D$ for a diffusive
system with the diffusion constant $D$ \cite{ChalkerIII,Prosenarx}. A scaling of the Thouless with $\ln N$ was 
also predicted for large single-particle Hilbert spaces in \cite{Chalker,ChalkerI} and for $s=1/2$ in \cite{ProsenI,ChalkerII} for  different systems by other methods.
By  the analysis in this section we generalize our results for the Thouless time valid in the perturbative regime of small $\Delta J$
 to arbitrary $\Delta J$.  


\section{Conclusion}

We studied the transition from ergodicity to localization in the KIC
of $N$ spins $1/2$ in the presence of strong local disorder. Our tool was
the double dual operator  introduced in \cite{Prosen} where it was
used  to prove ergodicity and RMT behavior of the
KIC with parameters  corresponding to the so called self-dual regime. We used it to
probe the localization effects caused by deviations from self-duality
considering that the behavior of the spectral form factor in the thermodynamic limit $N\gg1$
is determined by the largest eigenvalue $\lambda_{0}\left(  t\right)  $ of the double dual operator
$\mathcal{A}\left(  t\right)$.

We calculated $\lambda_{0}(t)$ as function of
the Ising constant $J$ in the interval from $\pi/4$ corresponding to
self-duality, to  $J=0$ corresponding to non-interacting spins, i.\ e.,
trivial localization.  Several methods were used which gave overlapping
result: First, for $J=0$ we obtained $\lambda_0(t)$ analytically and perturbatively for 
small $J$.  Comparing it with the numerically computed $\lambda_0(t)$ for $t\leq20$
we observe that despite a shrinking convergence of the perturbation theory with increasing 
$t$ it reproduces the structure observed in Ref.\ \cite{Braun} in the vicinity of $J=0$ where the 
$\lambda_0(t)$ values for a sufficiently large $J$ oscillate around the ones for $J=0$. We also
show that the eigenvalue $\lambda_0(t)$ dominating the spectral form factor for $J$ close to zero 
possesses the same symmetry as the dominant eigenvalue for small $t$ in the regime of $J$ close to $\pi/4$. 
Second, for times of the order of the Heisenberg time $T_{H}=2^{N}$ we estimated
$\lambda_{0}\left(  t\right)  $ by extensive numerical simulations computing the spectral form factor of
chains of up to $N\leq15$ and extrapolating the result to an infinite chain thereby complementing
our results in Ref.\ \cite{Braun} only valid for small $t$ and $N$, respectively. 

We observed here that the chain has two characteristic regimes: In the localized
regime the eigenvalue $\lambda_{0}(t)>1$ is separated by a gap from all other
eigenvalues of  the double dual operator; the form factor has an
exponential dependence on the number of spins. In the ergodic regime
$\lambda_{0}\left(  t\right)  $ is for large $t$ close to unity and is  the uppermost level
of  a narrow multiplet of $2t$ eigenvalues of $\mathcal{A}\left(  t\right) $ extending the picture of 
a single dominant largest eigenvalue in the regime of small $t$ we found in Ref.\ \cite{Braun}.
The form factor coincides with the RMT predictions and is
$N-$independent apart from the trivial scaling with $T_{H}$. Both regimes
coexist if the deviation  $\Delta J=\pi/4-J$ from self-duality is smaller than
$\Delta J_{c}\approx0.58$, namely we have localization with $\lambda
_{0}\left(  t\right)  >1$  when $t$ is smaller than the Thouless time and ergodicity with $\lambda
_{0}(t)\approx1$ when $t$ is larger than the Thouless time. The latter time
 grows from $0$ at $\Delta J=0$ to infinity at $\Delta
J=\Delta J_{c}$  and determines the transition in the 
thermodynamic limit. In the time interval below the Thouless time 
$\lambda_{0}\left(  t\right)  $ decays as a function of time, this decay
 tends to zero as $\Delta J$ approaches the critical value. 
For $\Delta J>\Delta J_{c}$ the eigenvalue
$\lambda_{0}$ is time independent apart from the system-specific fluctuations.

There are various possibilities to extend these results: First, our perturbative 
results will probably allow to get an analytical expression for the spectral Lyapunov
exponents in the symmetry block invariant under permutations and reflections in \cite{ChalkerI}. 
Second, our present findings
show that the behavior of the dominant eigenvalue $\lambda_{0}\left(
t\right)  $ of the  double dual operator  can be a useful
 localization witness in spin chains within the dual operator
approach. It would be interesting to check that for other spin values and
types of interactions.

\section*{Acknowledgements}
We acknowledge support by the Deutsche
Forschungsgemeinschaft through Project No.\ Gu431/9-1
(the “Dreiburg cooperation”).

\appendix

\section{Symmetry properties of eigenvector to the eigenvalue of largest magnitude for $J=0$}\label{appA}

According to Ref. \cite{Braun} the dual operator in the absence of interaction is
\begin{align}
\hat{W}\left(  t\right)    & =\prod_{n=1}^{N}\hat{w}_{n}\,,\\
\hat{w}_{n}  & =\hat{w}_{I}\hat{w}_{n,b},\nonumber\\
\left\langle \pmb{\sigma}^{t}\left\vert \hat{w}_{I}\right\vert
\pmb{\sigma}^{\prime t}\right\rangle  & =\left.\exp\left(  -iJ\sum_{k=1}%
^{t}\sigma_{k}\sigma_{k}^{\prime}\right)\right|_{J=0}=1  ,\nonumber\\
\left\langle \pmb{\sigma}^{t}\left\vert \hat{w}_{n,b}\right\vert
\pmb{\sigma}^{\prime t}\right\rangle  & =\delta_{\mathbf{\sigma}%
^{t}\mathbf{\sigma}}\exp\left(  -ih_{n}\sum_{k=1}^{t}\sigma_{k}\right)
\prod_{k=1}^{t}R_{\mathbf{\sigma}_{k}\sigma_{k+1}},\nonumber
\end{align}
i.\ e., the Ising matrix $\hat{w}_I$ has all elements equal 1. It has a single non-zero
eigenvalue equal to the matrix size $2^t$ corresponding to the eigenvector with
all elements equal, i.\ e., the sum of all basis vectors in the dual space with equal
coefficients; unnormalized it is
\begin{eqnarray}
|\chi\rangle&=&\sum_{\sigma_1,\sigma_2,\ldots,\sigma_t=\pm1}|\sigma_1,\sigma_2,\ldots,\sigma_t\rangle,\\
\langle\chi|\chi\rangle&=&2^t.\nonumber
\end{eqnarray}
Therefore $\hat{w}_I$ is simply,
\begin{equation}
\hat{w}_I=|\chi\rangle\langle\chi|
\end{equation}
That state is invariant with respect to the cyclic
shifts and reflections in the dual space the corresponding symmetry block we denoted in Ref.\ \cite{Braun} by (0+).
The local dual operator on the $n$-th spin is therefore,
\begin{equation}
\hat{w}_n=\sum_\sigma f_n(\pmb{\sigma})|\chi\rangle\langle\pmb{\sigma}|
\end{equation}
with 
\begin{equation}
f_n(\pmb{\sigma})=\langle\chi|\pmb{\sigma}\rangle\exp\left(-ih_n\sum_{k=1}^t\sigma_k\right)\prod_{k=1}^tR_{\sigma_k\sigma_{k+1}}.
\end{equation}
We know that the operators $\hat{w}_n$ are block diagonal in the symmetrized basis set.
However, $\hat{w}_n$ has only zero matrix elements in any symmetry block except (0+) 
because of the fully symmetric ket-vector $|\chi\rangle$. Hence all blocks of $\hat{w}_n$ except
(0+) are zero and same refers to their product $\hat{W}$. 
In the operator $\mathcal{A}$ the only non-zero block is therefore (0+; 0+) such that any non-zero eigenvalue
of $\mathcal{A}$ belongs to an eigenstate of highest symmetry.

\section{Perturbation theory}\label{appB}
Here we aim at obtaining a perturbative expression in $J$ for $\lambda_0(t)$. Therefore we expand the form factor (\ref{formfac})
perturbatively for small coupling $J$. In this context we need the linear and the quadratic orders in $J$ of ${\rm Tr}\hat{ U}^t$. The linear order is given by 
\begin{eqnarray}\label{1st}
&&t{\rm Tr}\left[\left(iJ\sum_{n=1}^N\sigma_n^z\sigma_{n+1}^z\right)\prod_{n=1}^N\hat{u}_n^{t}\right]\\
&=& t{\rm Tr}\left[\left(iJ\sum_{n=1}^N\sigma_n^z\sigma_{n+1}^z\right)\prod_{n=1}^N e^{it\gamma_n{\bf{e}}_n{\pmb{\sigma}}_n/2}\right]\nonumber\\
&=&-itJ 2^N\left[\sum_{n=1}^N\frac{\sin(t\gamma_n/2)}{\sqrt{2+\cot^2h_n}}\frac{\sin(t\gamma_{n+1}/2)}{\sqrt{2+\cot^2h_{n+1}}}\prod_{i=1,i\neq n,n+1}^N\cos(t\gamma_i/2)\right].\nonumber
\end{eqnarray}
The square roots in the last line result from the $z$-components of the normal vectors ${\bf{e}}_n$, ${\bf{e}}_{n+1}$ of the rotation matrix $\hat{u}_n$.
The corresponding contribution to the spectral form factor in linear order of $J$ that follows from the expression above is zero because (\ref{1st})  is purely imaginary. 

For the quadratic order two further contributions are of relevance. At first one of the $t$ $\hat{U}_I$'s in Eq.\ (\ref{evolup}) entering the ${\rm Tr}\hat{U}^t$
can be expanded at quadratic order in $J$ leading to the contribution to ${\rm Tr}\hat{U}^t$
\begin{equation}
-\frac{J^2t}{2}{\rm Tr}\left[\sum_{n,m=1}^N\sigma_n^z\sigma_{n+1}^z\sigma_m^z\sigma_{m+1}^z\prod_{n=1}^N\hat{u}_n^{t}.
\right]
\end{equation}
To proceed we need to distinguish in the last equation the different cases
\begin{enumerate}
\item $n$, $n+1$, $m$, $m+1$ all different
\item $n=m$
\item $n=m-1$, $n=m+1$.
\end{enumerate}
Summing up the contributions from all the cases we obtain the corresponding contribution to ${\rm Tr}\hat{U}^t$  
\begin{eqnarray}
&&-J^2t2^{N-1}\left[\sum_{n\neq m\neq m+1\neq n+1}\frac{\sin(t\gamma_n/2)}{\sqrt{2+\cot^2h_n}}
\frac{\sin(t\gamma_{n+1}/2)}{\sqrt{2+\cot^2h_{n+1}}}\frac{\sin(t\gamma_m/2)}{\sqrt{2+\cot^2h_m}}
\right.\nonumber\\ &&\left.\frac{\sin(t\gamma_{m+1}/2)}{\sqrt{2+\cot^2h_{m+1}}}
\prod_{i\neq n\neq m\neq n+1 \neq m+1}\cos(t\gamma_i/2)
+N\prod_{i=1}^N\cos(t\gamma_i/2)\right.\nonumber\\&&\left.-2\sum_n\frac{\sin(t\gamma_n/2)}{\sqrt{2+\cot^2h_n}}\frac{\sin(t\gamma_{n+2}/2)}{\sqrt{2+\cot^2h_{n+2}}}\prod_{i\neq n\neq n+2}\cos(t\gamma_i/2)
\right].
\end{eqnarray}
Furthermore we also need to expand two of the $t$ $\hat{U}_I$'s up linear order in $J$ yielding 
\begin{equation}
-\frac{J^2t}{2}\sum_{\tau=1}^{t-1}{\rm Tr} \left[\sum_{nm}\sigma_n^z\sigma_{n+1}^z\prod_i\hat{u}_i^{\tau}\sigma_m^z\sigma_{m+1}^z\prod_i\hat{u}_i^{t-\tau}\right]
\end{equation}
that can be transformed into
\begin{equation}
-\frac{J^2t}{2}\sum_{\tau=1}^{t-1}{\rm Tr} \left[\sum_{nm}\sigma_n^z\sigma_{n+1}^z\sigma_m^z\sigma_{m+1}^z\prod_{i\neq m\neq m+1}\hat{u}_i^{t}\hat{\overline{u}}_m^{\tau}\hat{\overline{u}}_{m+1}^{\tau}\hat{u}_m^{(t-\tau)}
\hat{u}_{m+1}^{(t-\tau)}\right],
\end{equation}
where $\hat{\overline{u}}_m^{\tau} =\cos(\tau\gamma_m/2)-(in_m^x\sigma_m^x+in_m^y\sigma_m^y-in_m^z\sigma_m^z)\sin(\tau\gamma_m/2)$ and an analogous definition of  $\hat{\overline{u}}_{m+1}^{\tau}$. Again distinguishing the cases
\begin{enumerate}
\item $n$, $n+1$, $m$, $m+1$ all different
\item $n=m$
\item $n=m-1$, $n=m+1$
\end{enumerate}
we get the following contribution to ${\rm Tr}\hat{U}^t$
\begin{eqnarray}
&&-J^2t2^{N-1}\sum_{\tau=1}^{t-1}\left[\sum_{n\neq m\neq m+1\neq n+1}\frac{\sin t\gamma_n/2}{\sqrt{2+\cot^2 h_n}}\frac{\sin t\gamma_{m+1}/2}{\sqrt{2+\cot^2 h_{m+1}}}\frac{\sin t\gamma_m/2}{\sqrt{2+\cot^2 h_m}}\frac{\sin t\gamma_{n+1}/2}{\sqrt{2+\cot^2 h_{n+1}}}\right.\nonumber\\&&\left.\prod_{i\neq n\neq m\neq n+1\neq m+1}\cos\frac{t\gamma_i}{2}+\sum_m\left(\cos\frac{\tau\gamma_m}{2}\cos\frac{(t-\tau)\gamma_m}{2}+
\sin\frac{\tau\gamma_m}{2}\right.\right.\nonumber\\&&\left.\left.\sin\frac{(t-\tau)\gamma_m}{2}\frac{\cot^2h_m}{2+\cot^2h_m}
\right)\left(\cos\frac{\tau\gamma_{m+1}}{2}\cos\frac{(t-\tau)\gamma_{m+1}}{2}+
\sin\frac{\tau\gamma_{m+1}}{2}\right.\right.\nonumber\\&&\left.\left.\sin\frac{(t-\tau)\gamma_{m+1}}{2}\frac{\cot^2h_{m+1}}{2+\cot^2h_{m+1}}
\right)\prod_{i\neq m,m+1}\cos\frac{t\gamma_i(h)}{2}-\sum_m\left(\cos\frac{\tau\gamma_m}{2}\cos\frac{(t-\tau)\gamma_m}{2}\right.\right.\nonumber\\&&\left.\left.+
\sin\frac{\tau\gamma_m}{2}\sin\frac{(t-\tau)\gamma_m}{2}\frac{\cot^2h_m}{2+\cot^2h_m}
\right)\frac{\sin t\gamma_{m-1}/2}{\sqrt{2+\cot^2 h_{m-1}}}\frac{\sin t\gamma_{m+1}/2}{\sqrt{2+\cot^2 h_{m+1}}}\right.\nonumber\\&&\left.\prod_{i\neq m\neq m-1\neq m+1}\cos\frac{t\gamma_i}{2}
-\sum_m\left(\cos\frac{\tau\gamma_{m+1}}{2}\cos\frac{(t-\tau)\gamma_{m+1}}{2}\right.\right.\nonumber\\&&\left.\left.+
\sin\frac{\tau\gamma_{m+1}}{2}\sin\frac{(t-\tau)\gamma_{m+1}}{2}\frac{\cot^2h_{m+1}}{2+\cot^2h_{m+1}}
\right)\frac{\sin t\gamma_{m}/2}{\sqrt{2+\cot^2 h_{m}}}\frac{\sin t\gamma_{m+2}/2}{\sqrt{2+\cot^2 h_{m+2}}}\right.\nonumber\\&&\left.\prod_{i\neq m\neq m+1\neq m+2}\cos\frac{t\gamma_i}{2}
\right].
\end{eqnarray}
Now we can obtain the contribution to the spectral form factor that is of quadratic order in $J$. To summarize, we just derived an expansion of ${\rm Tr} \hat{U}^t$ in $J$
\begin{equation}
{\rm Tr}\hat{U}^t=\left({\rm Tr} \hat{U}^t\right)_{(0)}+J\left({\rm Tr} \hat{U}^t\right)_{(1)}+J^2\left({\rm Tr} \hat{U}^t\right)_{(2)}+\dots,
\end{equation}
where the subscripts in the last equation indicate the order in $J$. 
This allows to obtain the contribution to $K_N(t)$ quadratic in $J$
\begin{equation}
\left({\rm Tr} \hat{U}^t\right)_{(1)}\left({\rm Tr} \hat{U}^t\right)_{(1)}^*+\left({\rm Tr} \hat{U}^t\right)_{(2)}\left({\rm Tr} \hat{U}^t\right)_{(0)}^*+\left({\rm Tr} \hat{U}^t\right)_{(0)}\left({\rm Tr} \hat{U}^t\right)_{(2)}^*.
 \end{equation}
Summing all these terms we get for the contribution quadratic in $J$ to $K_N(t)$ 
\begin{eqnarray}\label{eq10}
&&t^2J^24^N\left[\sum_n\frac{\sin^2t\gamma_n/2}{2+\cot^2h_n}\frac{\sin^2t\gamma_{n+1}/2}{2+\cot^2h_{n+1}}
\prod_{i\neq n\neq n+1}\cos^2\frac{t\gamma_i}{2}\right.\nonumber\\&&\left.+2\sum_n\frac{\sin t\gamma_n/2}{\sqrt{2+\cot^2h_n}}\frac{\sin^2 t\gamma_{n+1}/2}{2+\cot^2h_{n+1}}\frac{\sin t\gamma_{n+2}/2}{\sqrt{2+\cot^2h_{n+2}}}\prod_{i\neq n\neq n+1\neq n+2}\right.\nonumber\\&&\left.\cos^2\frac{t\gamma_i}{2}\cos\frac{t\gamma_n}{2}
\cos\frac{t\gamma_{n+2}}{2}-\frac{N}{T}\prod_i\cos^2\frac{t\gamma_i}{2}+\frac{2}{t}\right.\nonumber\\&&\left.\sum_n
\frac{\sin t\gamma_n/2}{\sqrt{2+\cot^2h_n}} \frac{\sin t\gamma_{n+2}/2}{\sqrt{2+\cot^2h_{n+2}}}
\prod_{i\neq n\neq n+2}\cos^2\frac{t\gamma_i}{2}\cos\frac{t\gamma_n}{2}\cos\frac{t\gamma_{n+2}}{2}
\right.\nonumber\\&&\left. -\frac{1}{t}\sum_{\tau=1}^{t-1}\sum_n\left(\cos\frac{\tau\gamma_n}{2}\cos\frac{(t-\tau)\gamma_n}{2}+\sin\frac{\tau\gamma_n}{2}\sin \frac{(t-\tau)\gamma_n}{2}\frac{\cot^2h_n}{2+\cot^2h_n}\right)
\right.\nonumber\\&&\left.
\left(\cos\frac{\tau\gamma_{n+1}}{2}\cos\frac{(t-\tau)\gamma_{n+1}}{2}+\sin\frac{\tau\gamma_{n+1}}{2}\sin \frac{(t-\tau)\gamma_{n+1}}{2}\frac{\cot^2h_{n+1}}{2+\cot^2h_{n+1}}\right)\right.\nonumber\\&&\left.\prod_{i\neq n\neq n+1}\cos^2\frac{t\gamma_i}{2}\cos\frac{t\gamma_n}{2}\cos\frac{t\gamma_{n+1}}{2}+\frac{2}{t}\sum_n
\left(\cos\frac{\tau\gamma_n}{2}\cos\frac{(t-\tau)\gamma_n}{2}\right.\right.\nonumber\\&&\left.\left.+\sin\frac{\tau\gamma_n}{2}\sin \frac{(t-\tau)\gamma_n}{2}\frac{\cot^2h_n}{2+\cot^2h_n}\right)\frac{\sin t\gamma_{n-1}/2}{\sqrt{2+\cot^2 h_{n-1}}}
\frac{\sin t\gamma_{n+1}/2}{\sqrt{2+\cot^2 h_{n+1}}}\right.\nonumber\\&&\left.\prod_{i\neq n\neq n-1\neq n+1}\cos^2\frac{t\gamma_i}{2}
\cos\frac{t\gamma_n}{2}\cos\frac{t\gamma_{n-1}}{2}\cos\frac{t\gamma_{n+1}}{2}
\right].
\end{eqnarray}
Using the relation
\begin{equation}
\int_0^\pi dh\cos\frac{t\gamma_{i}}{2}\sin\frac{t\gamma_{i}}{2}=0
\end{equation}
with $\gamma_i$ given in Eq.\ (\ref{gamma}) and the result in Eq.\ (\ref{noninteract})
we can simplify (\ref{eq10}) to
\begin{eqnarray}\label{eq11}
&&tJ^2N\left[t\left(\frac{4}{\pi}\int_0^\pi dh\frac{\sin^2t\gamma/2}{2+\cot^2h}\right)^2
\left(2+\frac{(-1)^q(2p-1)!!}{(2p)!!}\right)^{N-2}\right.\nonumber\\&&\left.
-\left(2+\frac{(-1)^q(2p-1)!!}{(2p)!!}\right)^{N}
\right.\nonumber\\&&\left. -\sum_{\tau=1}^{t-1}\left(\frac{4}{\pi}\int_0^\pi dh\left(\cos\frac{\tau\gamma}{2}\cos\frac{(t-\tau)\gamma}{2}+\sin\frac{\tau\gamma}{2}\sin \frac{(t-\tau)\gamma}{2}\frac{\cot^2h}{2+\cot^2h}\right)
\right.\right.\nonumber\\&&\left.\left.\cos\frac{t\gamma}{2}\right)^2
\left(2+\frac{(-1)^q(2p-1)!!}{(2p)!!}\right)^{N-2}
\right].
\end{eqnarray}
with $q=[(t+1)/2]$ and $p=[t/2]$. This implies we can rewrite the spectral form factor as 
\begin{equation}
K_N(t)=\lambda_0(t)^N\left[1+\frac{NJ^2f(t)}{\lambda_0^2}\right]
\end{equation}
with a function $f(t)$ that can be read off from Eq.\ (\ref{eq11}) and 
\begin{equation}
\lambda_0(t)=\left(2+\frac{(-1)^q(2p-1)!!}{(2p)!!}\right).
\end{equation}
Using $1+Nx\approx(1+x)^N$ in the limit $N\to\infty$ we get for the spectral form factor
\begin{equation}
K_N(t)=\left[\lambda_0(t)\left(1+\frac{J^2f(T)}{\lambda_0(t)^2}\right)\right]^N.
\end{equation}
From this equation we can extract by means of $K_N(t)=\lambda(t)^N$ an expression for $\lambda(t)$
\begin{equation}
\lambda(t)=\lambda_0(t)\left(1+\frac{J^2f(t)}{\lambda_0(t)^2}\right)
\end{equation}
that yields in detail
\begin{eqnarray}\label{eq100}
\lambda(t)&=&\lambda_0(t)+\frac{tJ^2}{\lambda_0(t)}\left[T\left(\frac{4}{\pi}\int_0^\pi dh\frac{\sin^2t\gamma/2}{2+\cot^2h}\right)^2
-\lambda_0(t)^2
\right.\nonumber\\&&\left. -\sum_{\tau=1}^{t-1}\left(\frac{4}{\pi}\int_0^\pi dh\left(\cos\frac{\tau\gamma}{2}\cos\frac{(t-\tau)\gamma}{2}+\sin\frac{\tau\gamma}{2}\sin \frac{(t-\tau)\gamma}{2}\frac{\cot^2h}{2+\cot^2h}\right)
\right.\right.\nonumber\\&&\left.\left.\cos\frac{t\gamma}{2}\right)^2
\right].
\end{eqnarray}
These integrals can be calculated by residual integration in terms of the generalized binomial coefficients. The final result is given as expression (\ref{final}) in the 
main text.

\begin{figure}\begin{center}
\includegraphics[width=10cm]{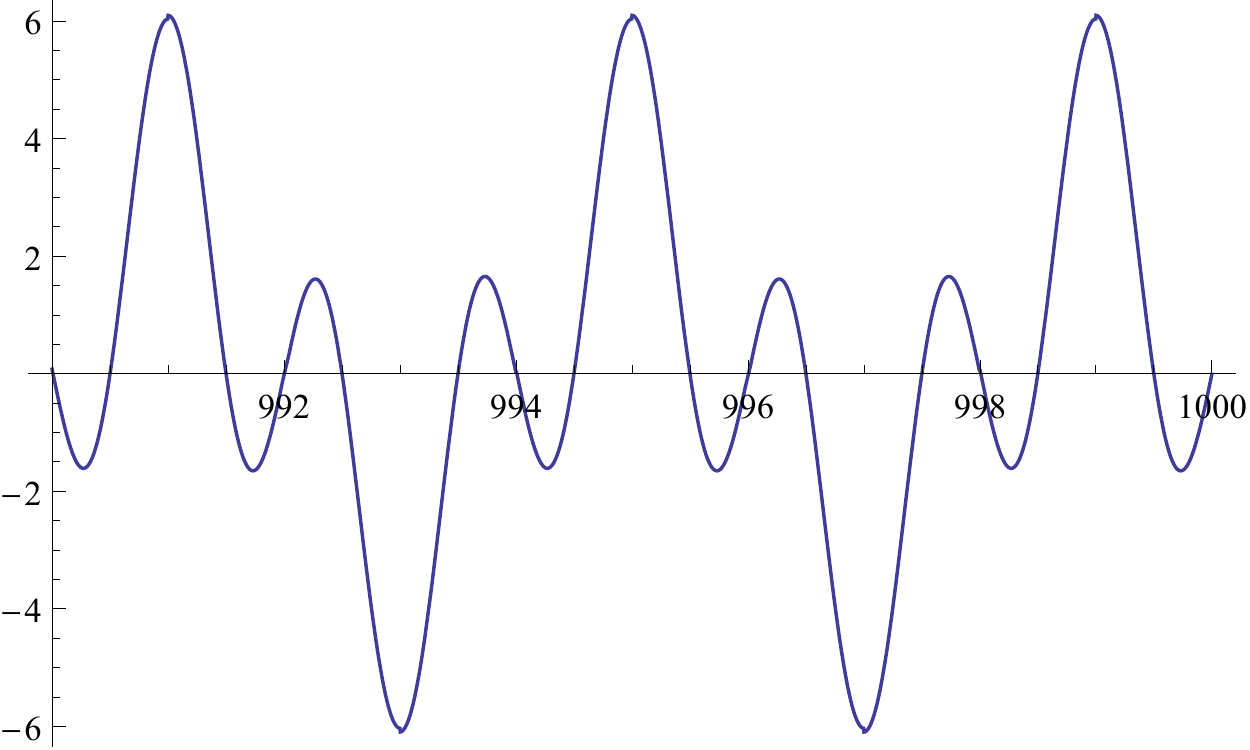}
\caption{Contribution (\ref{eqq30}) as a function of $t$.}
\label{osc}
\end{center}\end{figure}

The expression (\ref{final}) still contains one sum with respect to $\tau$ that we now want to perform in the limit of large $t$.
We will concentrate on the smooth part ignoring the contributions from oscillations with period four. In this context we use the method of stationary phase assuming $\tau$, $t$, and $t-\tau$ to be large. We note that this is in principle not fulfilled as not for all terms in the $\tau$-sum $\tau$ and $t-\tau$ can be large, but we found by comparison with numerics that our approximation is already quite good for $t\approx20$. In a first step we rewrite 
\begin{eqnarray}\label{eq999}
&&\sum_{\tau=1}^{t-1}\left(\frac{4}{\pi}\int_0^\pi dh\left(\cos\frac{\tau\gamma(h)}{2}\cos\frac{(t-\tau)\gamma(h)}{2}+\sin\frac{\tau\gamma(h)}{2}\sin \frac{(t-\tau)\gamma(h)}{2}\frac{\cot^2h}{2+\cot^2h}\right)
\cos\frac{t\gamma(h)}{2}\right)^2\nonumber\\
&=&\sum_{\tau=1}^{t-1}\frac{1}{\pi^2}\left[\int_0^\pi dh\left(\cos t\gamma(h)+1\right)\frac{2}{2+\cot^2h}+\left(\cos\tau \gamma(h)+\cos(t-\tau)\gamma(h)\right)\frac{2+2\cot^2h}{2+\cot^2h}\right]^2.
\end{eqnarray}
 Taking into account that $\gamma(h)$ becomes stationary at $h=0$ and at $h=\pi$ we get for the last expression
\begin{eqnarray}\label{eq1001}
&&\sum_{\tau=1}^{t-1}\frac{1}{\pi}\left[\sqrt{\frac{1}{\tau}}\left(\cos\left(\frac{\tau\pi}{2}+\frac{\pi}{4}\right)+\cos\left(\frac{3\tau\pi}{2}-\frac{\pi}{4}\right)\right)+\sqrt{\frac{1}{t-\tau}}\left(\cos\left(\frac{(t-\tau)\pi}{2}+\frac{\pi}{4}\right)\right.\right.\nonumber\\&&\left.\left.+\cos\left(\frac{3(t-\tau)\pi}{2}-\frac{\pi}{4}\right)\right)
+\sqrt{\pi}(2-\sqrt{2})\right]^2,
\end{eqnarray}
where we evaluated the part of the integral in (\ref{eq999}) independent of $t$ and $\tau$ exactly and the rest within the stationary phase approximation. 
In the next step we want to evaluate the latter expression in the limit of large $t$ neglecting additionally contributions rapidly oscillating as a function of $t$. Calculating the square in (\ref{eq1001}) we get the following contributions: First, 
\begin{equation}\label{eqc}
(2-\sqrt{2})^2(t-1).
\end{equation}  
Second, the mixed term from the square in (\ref{eq1001}) yields 
\begin{eqnarray}
&&\sum_{\tau=1}^{t-1}\frac{4(2-\sqrt{2})}{\sqrt{\pi \tau}}\left[\cos\left(\frac{\tau\pi}{2}+\frac{\pi}{4}\right)+\cos\left(\frac{3\pi \tau}{2}-\frac{\pi}{4}\right)\right]\nonumber\\&&=\frac{4(\sqrt{2}-1)}{\sqrt{\pi}}\left[(2 - \sqrt{2}) \zeta[1/2] - 
   i^t \left(\zeta[1/2, t/4] - \zeta[1/2, (t+2)/4]\right) \right.\\&&\left.-\zeta[1/2, 1/4] + \zeta[1/2, 3/4] - 
 i^t \left(\zeta[1/2, 1/4 (3 + t)] - \zeta[1/2, (1 + t)/4]\right)\right]\nonumber
\end{eqnarray}
for $t$ even and for $t$ odd
\begin{eqnarray}
&&\sum_{\tau=1}^{t-1}\frac{4(2-\sqrt{2})}{\sqrt{\pi \tau}}\left[\cos\left(\frac{\tau\pi}{2}+\frac{\pi}{4}\right)+\cos\left(\frac{3\pi \tau}{2}-\frac{\pi}{4}\right)\right]\nonumber\\&&=\frac{4(\sqrt{2}-1)}{\sqrt{\pi}}\left[(2 - \sqrt{2}) \zeta[1/2] +\sin[\pi t/2] 
 \left(\zeta[1/2, (1+t)/4] - \zeta[1/2, (3 + t)/4]\right) \right.\\&&\left.-\zeta[1/2, 1/4] + \zeta[1/2, 3/4] +(-1)^{[t/2]} 
  \left(\zeta[1/2, 1/4 (1 +2[ t/2])] - \zeta[1/2, (3 + 2[t/2])/4]\right)\right],\nonumber
\end{eqnarray}
where the last lines in the two last equations are obtained from mathematica with the Hurwitz zeta function $\zeta[a,s]$ and the Riemann zeta function $\zeta[s]$.
As these functions tend to zero for $t\to\infty$ we get for the last expression in that limit
\begin{eqnarray}\label{eqc1}
\frac{4(\sqrt{2}-1)}{\sqrt{\pi}}\left[(2-\sqrt{2})\zeta[1/2]-\zeta[1/2,1/4]+\zeta[1/2,3/4]\right].
\end{eqnarray}
Third, the last contributions results from the square of the cosine functions in (\ref{eq1001}). Neglecting all contributions that decay with $t$ we get by mathematica
\begin{equation}\label{eqc2}
\sum_{\tau=1}^{t-1}\frac{2}{\pi \tau}\left[\cos\left(\frac{\tau\pi}{2}-\frac{\pi}{4}\right)+\cos\left(\frac{3\pi \tau}{2}-\frac{\pi}{4}\right)\right]^2=\frac{4}{\pi}\left(\gamma+\ln t\right)
\end{equation} 
with the Euler constant $\gamma$. The mixed term 
\begin{eqnarray}\label{eqq30}
&&\sum_{\tau=1}^{t-1}\frac{2}{\pi \sqrt{\tau(t-\tau)}}\left[\cos\left(\frac{\tau\pi}{2}-\frac{\pi}{4}\right)+\cos\left(\frac{3\pi \tau}{2}-\frac{\pi}{4}\right)\right]\nonumber\\ &&\left[\cos\left(\frac{(t-\tau)\pi}{2}-\frac{\pi}{4}\right)+\cos\left(\frac{3\pi(t- \tau)}{2}-\frac{\pi}{4}\right)\right]
\end{eqnarray}
oscillates around zero as a function of $t$, see Fig.\ \ref{osc} and thus does not contribute.

To obtain an overall expression for Eq.\ (\ref{eq100}) averaged with respect to $t$ we  also need to
analyze the first term there in the curved bracket. In the limit of large $t$ we get
\begin{equation}\label{eqsin}
\frac{4}{\pi}\int_0^\pi dh\frac{\sin^2t\gamma/2}{2+\cot^2h}=2-\sqrt{2}.
\end{equation}  
To summarize, the results in Eqs.\ (\ref{eqsin},\ref{eqc2},\ref{eqc1},\ref{eqc},\ref{lamunp}) finally imply the relation (\ref{final1}) in the main part.

\section{Fitting by a hyperbola}\label{hyperbola}

\begin{figure}[h]
 \begin{center}
  \includegraphics[width=10cm]{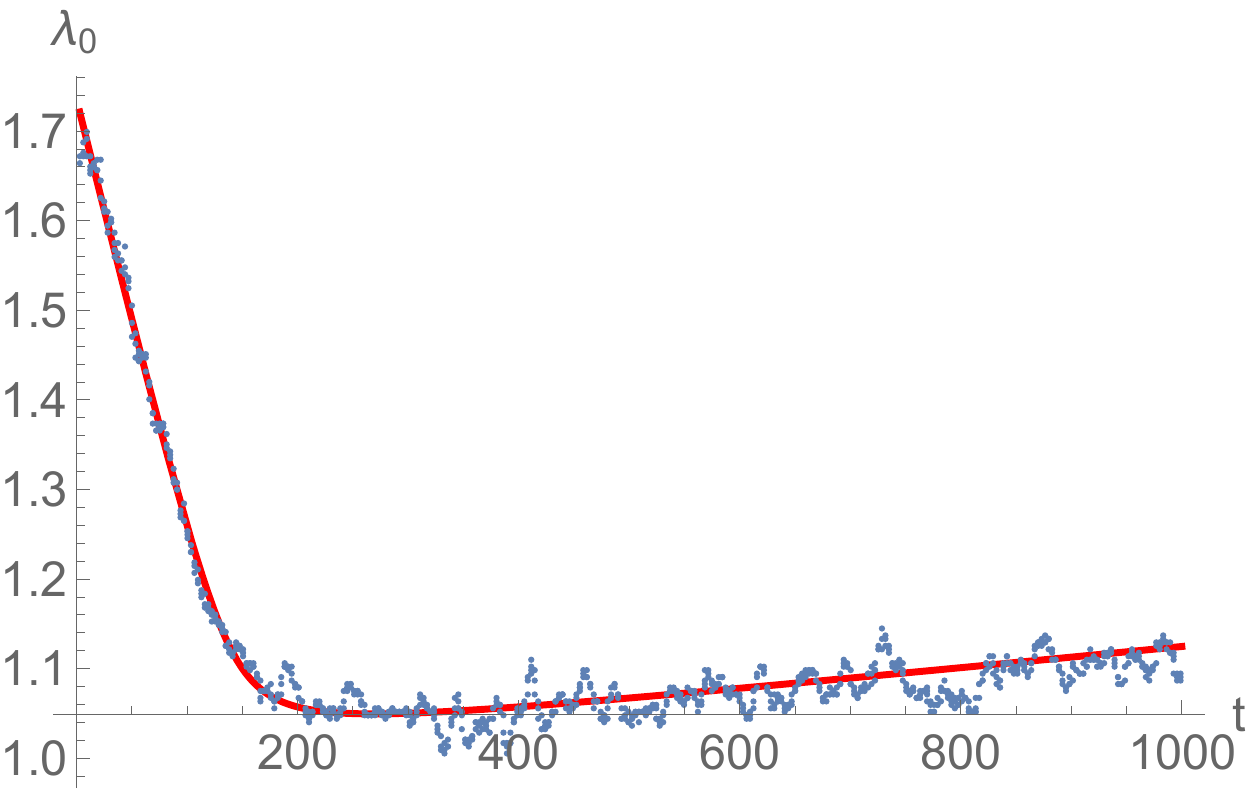}
\caption{$\lambda_{0,13}$ against time (blue dots), fitting by a hyperbola (red line)
}
  \label{Fig6}
 \end{center}
\end{figure}

The empirical curves of $\lambda_{0,N}\left(  t\right)  $ remind of a hyperbola
approaching straight lines on both edges and a curved arc between them.
Therefore we used for fitting a 3-parameter function representing a hyperbola with
one of whose asymptotes coinciding with the GOE prediction $\lambda
_{0,N}^{GOE}\left(  t\right)  $ (\ref{GOElambda}),
\begin{align*}
\lambda_{fit,N}\left(  t\right)    & =-b-2at\\
& +\sqrt{b^{2}-4ac+2\left(  1+b\right)  \left(  2a+2^{-N}\right)  t+\left(
2a+2^{-N}\right)  ^{2}t^{2}}%
\end{align*}
with $a,b,c$ the fitting parameters. That fitting worked well for all $\Delta
J$. As an example see Fig.\ \ref{Fig6} where we show $\lambda_{0,13}\left(  t\right)
,\Delta J=0.4,$ (blue dots) and the fitting curve (red line).%

\begin{thebibliography}{11}
\bibitem{footn} In fact $2t+1$, plus some number-theoretical exceptions, but such details
will be of not interest in our crude approach.
\end{thebibliography}
\end{document}